\documentclass{article} \usepackage{romp32,amsmath}

\begin{document}

\title{A Poisson Bracket on Multisymplectic Phase Space\thanks{%
\thinspace \thinspace \thinspace Work supported by FAPESP (Funda\c{c}\~{a}o
de Amparo \`{a} Pesquisa do Estado de S\~{a}o Paulo)}}
\author{Michael Forger \\
Departamento de Matem\'{a}tica Aplicada,\\
Instituto de Matem\'{a}tica e Estat\'{\i}stica \\
Universidade de S\~{a}o Paulo \\
Caixa Postal 66281 \\
BR-05315-970 S\~{a}o Paulo, S.P., Brazil \\
forger@ime.usp.br \\
[2ex] Hartmann R\"{o}mer\thanks{%
\thinspace\ Talk given by H. R\"{o}mer} \\
Fakult\"{a}t f\"{u}r Physik der Univ. Freiburg \\
Hermann-Herder-Str.3 \\
D-79104 Freiburg, Germany \\
roemer@physik.uni-freiburg.de}
\maketitle

\begin{abstract}
A new Poisson bracket for Hamiltonian forms on the full multisymplectic
phase space is defined. At least for forms of degree $n-1$, where $n$ is the
dimension of space-time, Jacobi's identity is fulfilled.
\end{abstract}

\bigskip


\section{Introduction}

The main idea of the multisymplectic formulation of classical field theory
defined by a Lagrangian density ${\cal L}$ consists in treating space and
time derivatives of fields on an equal footing. The advantage of this
approach, as compared to the common canonical formulation, is two-fold:

\begin{itemize}
\item Lorentz covariance is manifest and automatic.

\item Phase space is finite dimensional.
\end{itemize}

Also two-fold are the disadvantages:

\begin{itemize}
\item The introduction of several ``conjugate momenta'' $\, \pi_i^\mu =
\partial {\cal L} / \partial \, \partial _\mu \varphi ^i \,$ associated  to
every field component $\varphi^i$ destroys the usual duality between  fields
and momenta.

\item Quantization is unclear in the multisymplectic formalism.
\end{itemize}

A first step towards multisymplectic quantization is the formulation of
multisymplectic Poisson brackets. Pioneer work in this direction has been
done by Kijowski \cite{kijow} (see also \cite{kijtul}). We \cite{forroe}
were motivated by recent innovative work of Kanatchikov \cite{kanat} that
will be briefly sketched below. For a general comprehensive presentation of
the multisymplectic formalism, including references to the early literature
on the beginnings of the subject, which date back to the second decade of
the $20^{{\rm th}}$ century, see \cite{marsden}.

\bigskip

In terms of the covariant De Donder-Weyl Hamiltonian,
\begin{equation}
{\cal H}~=~\pi_i^\mu \, \partial_\mu \varphi^i - \, {\cal L}~,
\end{equation}
the equations of motion can be brought into the form
\begin{eqnarray}
&{\displaystyle \frac{\partial {\cal H}}{\partial \pi_i^\mu}}~
=~\partial_\mu \varphi^i~,& \\
&{\displaystyle \frac{\partial {\cal H}}{\partial \varphi^i}}~ =~- \,
\partial_\mu \pi_i^\mu~.
\end{eqnarray}
The geometry of the multisymplectic phase space $P$ can briefly be described
as follows: Let the field $\varphi$ be a section of a fibre bundle $\; E
\overset{\pi}{\longrightarrow} M \;$ over an $n$-dimensional space-time
manifold $M$ with fibre dimension $N$. Let $(x^\mu) _{\mu = 1...n}$ be local
coordinates on $M$ and $(q^i) _{i = 1...N}$ be local coordinates on the
standard fibre. The (first) jet bundle $J(E)$ of $E$ is an affine bundle of
fibre dimension $nN$ over $E$ and a bundle without special structure and
fibre dimension $nN+N$ over $M$. Local coordinates for $J(E)$ can be written
as
\begin{equation}
(x^\mu,q^i,q_\mu^i)~,
\end{equation}
where it is understood that the coordinates of the (first) jet of a section $%
\varphi$ of $E$ at the point $x$ are given by
\begin{equation}
x^\mu~~~,~~~q^i~=~\varphi^i(x)~~~,~~~q_\mu^i~=~\partial_\mu \varphi^i (x)~.
\end{equation}
The multisymplectic phase space $P$ is given by the total space of the
(first) cojet bundle $J^*(E)$ of fibrewise affine mappings
\begin{equation}
J(E)~\longrightarrow~\pi_E^* \left( \Lambda^n \, T^\ast M \right)
\end{equation}
into $n$-forms over $M$. $J^*(E)$ is a vector bundle of fibre dimension $nN+1
$ over $E$. Representing such a fibrewise affine mapping in the form $\;
q_\mu^i \longmapsto \left( p_i^\mu q_\mu^i + p \right) d^n x$, local
coordinates for $J^*(E)$ can be written as
\begin{equation}
(x^\mu,q^i,p_i^\mu,p)~.
\end{equation}
The dimension of the multisymplectic phase space is
\begin{equation}
\dim P=(N+1)(n+1)~.
\end{equation}
Of fundamental importance are the canonical $n$-form $\theta$ and the
multisymplectic $(n+1)$-form $\omega$ on $J^*(E)$: these can be defined
intrinsically and have the coordinate expression
\begin{eqnarray}
&\theta~=~p_i^\mu \; dq^{i} \wedge d^n x_\mu \, - \, p \; d^n x~,& \\[1mm]
&\omega~=~- \, d\theta~=~dq^i \wedge dp_i^\mu \wedge d^n x_\mu \, + \, dp
\,\wedge d^n x~=~\omega^V \, + \, dp \,\wedge d^n x~,&
\end{eqnarray}
where $\, d^n x_\mu = i_{\partial_\mu} d^n x \,$ arises by contraction of
the volume element $d^n x$ with $\partial_\mu$ \cite{marsden}; $\omega^V$ is
an abbreviation for the ``vertical part'' of $\omega$ (which, in contrast to
$\omega$ itself, has no intrinsic meaning).

$P$ is a field theoretic generalization of the doubly extended phase space
of ordinary mechanics, whereas the submanifold
\begin{equation}
P_{{\cal H}}~=~\left\{ z\in P \; / \; p = {\cal H}(z) \right\}~,
\end{equation}
carrying forms $\theta_{{\cal H}}$, $\omega_{{\cal H}}$ and $\omega_{{\cal H}%
}^V$ obtained by restriction from the forms $\theta$, $\omega$ and $\omega^V$
on $P$, respectively, generalizes the extended phase space of mechanics:
this is the space used in Kanatchikov's approach \cite{kanat}. First, the
direct generalization of the Hamiltonian vector fields of classical
mechanics associated with given Hamiltonians are $n$-multivector fields $X_{%
{\cal H}}$ such that
\begin{equation}
i_{X_{{\cal H}}} \, \omega_{{\cal H}}^V~=~d^V {\cal H}~,
\end{equation}
where
\begin{equation}
d^V~:=~dq^i \wedge \frac{\partial}{\partial q^i} \, + \, dp_i^\mu \wedge
\frac{\partial}{\partial p_i^\mu}~.
\end{equation}
Moreover, Kanatchikov defines {\em Hamiltonian forms} of degree $p$ and {\em %
Hamiltonian multivector fields} of degree $n-p$ to be $p$-forms $F$ and $%
(n-p)$-multivector fields $X$ that can be related through the formula
\begin{equation}  \label{hamkan}
i_X \, \omega_{{\cal H}}^V~=~d^V F~,
\end{equation}
with the additional restriction that the form $F$ should be {\em horizontal}%
. Of course, when \linebreak $p > 0$, not every $p$-form is Hamiltonian
because equation (\ref{hamkan}) imposes a strong integrability condition on $%
F$. Finally, in analogy with classical mechanics, Kanatchikov defines a
generalized Poisson bracket between Hamiltonian forms of arbitrary degree by
\begin{equation}  \label{kanklammer}
\left\{ \overset{p}{F}_{1} , \overset{q}{F}_{1} \right\}~ =~(-1)^{n-p} \,
i_{X_{F_{1}}} i_{X_{F_{2}}} \omega _{{\cal H}}^V~.
\end{equation}
This bracket is well defined because $F$ determines $X_F$ up to an element
in the kernel of $\omega _{{\cal H}}^V$. Moreover, it can be checked that
Jacobi's identity is fulfilled.

The approach of Kanatchikov provides an important step forward in
multisymplectic dynamics, but it suffers from two evident shortcomings.

\begin{itemize}
\item The restriction to $P_{{\cal H}}$ and horizontal/vertical splitting do
a  great deal of violence to the multisymplectic structure and introduce
non-generic features like $d^V$ (which may be the price for formulating
dynamics in a Poisson-Hamiltonian framework).

\item The assumption of horizontality of Hamiltonian forms is too
restrictive,  as can be seen by considering the multimomentum map \cite%
{marsden} which  provides $(n-1)$-forms associated to generators of
symmetries (Noether  currents). Horizontality excludes symmetries associated
with nontrivial  transformations of space-time and an adequate treatment of
the  energy-momentum tensor.
\end{itemize}

As far as the first point is concerned, the situation has been alleviated by
recent work of Paufler \cite{paufler} who has shown that a vertical exterior
derivative $d^V$ can always be defined and that the bracket (\ref{kanklammer}%
) does not depend on the ambiguities inherent in its definition.

Our proposal \cite{forroe} is to completely avoid all these problems by
working directly on the full multisymplectic phase space $\, P = J^*(E)$.
Hamiltonian forms $f$ and Hamiltonian multivector fields $X_f$ are defined
on $P$, without any horizontality restriction on $f$, and are related by
means of the full multisymplectic form $\omega$, according to
\begin{equation}
i_{X_f} \, \omega~=~df  \label{eq:hamformvf}
\end{equation}

\section{Hamiltonian multivector fields and forms}

The Lie derivative of differential forms along vector fields can be
generalized to a Lie derivative of differential forms along multivector
fields, defined as the graded commutator between the exterior derivative $d$
and the respective contraction operator: for a $p$-multivector field $X$ on $%
P$,
\begin{equation}
L_{X}\,\alpha ~=~\left[ \,d,i_{X}\right] \alpha ~=~\Bigl(d\,i_{X}-(-1)^{p}%
\,i_{X}\,d\Bigr)\alpha ~.  \label{liederivative}
\end{equation}%
On the other hand, we have the Schouten bracket between multivector fields,
which is the (unique) extension of the Lie bracket between vector fields by
graded derivations (provided one uses an appropriately shifted degree).
These operations satisfy the following relations:
\begin{eqnarray}
&&\left[ d,L_{X}\right] ~=~L_{X}\,-\,(-1)^{p-1}L_{X}\,d~=~0~, \\[1mm]
&&i_{\left[ X,Y\right] }\alpha ~=~(-1)^{(p-1)q}\left(
L_{X}\,i_{Y}\,-\,(-1)^{(p-1)q}i_{Y}\,L_{X}\right) \alpha   \label{schouten}
\\[1mm]
&&L_{\left[ X,Y\right] }\alpha ~=~(-1)^{(p-1)(q-1)}\left(
L_{X}L_{Y}\,-\,(-1)^{(p-1)(q-1)}L_{Y}L_{X}\right) \alpha
\end{eqnarray}%
A multivector field $X$ on $P$ is called {\em locally Hamiltonian} or {\em %
multisymplectic} if
\begin{equation}
L_{X}\,\omega ~=~0~.
\end{equation}%
A direct consequence of equation (\ref{liederivative}) is \\[3mm]
{\bf Lemma:} {\em Every Hamiltonian multivector field is multisymplectic.} \\%
[3mm]
From equation (\ref{schouten}) we readily infer \\[3mm]
{\bf Lemma:} {\em The Schouten bracket $[X,Y]$ of two multisymplectic
multivector fields $X$ and $Y$ is Hamiltonian.} \\[3mm]
Proof:
\begin{eqnarray*}
i_{\left[ X,Y\right] }\,\omega \!\! &=&\!\!\pm \left[ i_{Y},L_{X}\right]
\,\omega ~=~\pm \,L_{X}\,i_{Y}\,\omega  \\
&=&\!\!\pm \,i_{X}\,d\,i_{Y}\,\omega \pm d\,i_{X}\,d\,i_{Y}\,\omega  \\
&=&\!\!\pm \,d\,i_{X}\,i_{Y}\,\omega
\end{eqnarray*}

In what follows, we shall consider mainly vector fields, rather than the
more general multivector fields, on $M$, $E$, $J(E)$ and $J^{\ast }(E)$. In
particular, Hamiltonian vector fields $X$ on $J^{\ast }(E)$ will play a
prominent role. Their associated Hamiltonian forms $f$ are of degree $n-1$,
and the Poisson bracket of two Hamiltonian $(n-1)$-forms will again be a
Hamiltonian $(n-1)$-form. (More generally, the Poisson bracket of a
Hamiltonian $(n-1)$-form with a Hamiltonian $p$-form will again be a
Hamiltonian $p$-form.)

We begin with vector fields $X_M$ on $M$ and vector fields $X_E$ on $E$:
they generate diffeomorphisms of $M$ and of $E$, respectively. $E$ being not
just any manifold but the total space of a fibre bundle $\; E \overset{\pi}{%
\longrightarrow} M$, there are two special classes of vector fields on $E$,
namely {\em projectable vector fields} that generate bundle automorphisms of
$E$ (covering diffeomorphisms of $M$) and {\em vertical vector fields} that
generate strict bundle automorphisms of $E$ (covering the identity on $M$).
By definition, a vector field $X_E$ on $E$ is projectable (or more
precisely, $M$-projectable) iff there exists a vector field $X_M$ on $M$
such that
\begin{equation}
T \pi \, X_E (e)~=~X_M(\pi(e))
\end{equation}%
for all $\, e \in E$, and is vertical if this formula holds with $\, X_M = 0$%
. In local coordinates $(x^\mu)$ on $M$ and $(x^\mu,q^i)$ on $E$, writing
\begin{eqnarray}
&X_M~=~X^\mu {\displaystyle \frac{\partial}{\partial x^\mu}}~, \\
&X_E~=~X^\mu {\displaystyle \frac{\partial}{\partial x^\mu}} \, + \, X^i {%
\displaystyle \frac{\partial}{ \partial q^i}}~,
\end{eqnarray}
we see that $X_E$ is projectable iff the $X^\mu$ are independent of the
fibre coordinates $q^i$ and that $X_E$ is vertical iff the $X^\mu$ vanish.
Now the jet bundles $J(E)$ and the cojet bundle $J^*(E)$ are bundles over $E$
for which we have the following \\[3mm]
{\bf Theorem:} {\em Bundle automorphisms $\Phi_E$ of $E$ over $M$ can be
lifted to bundle automorphisms $\Phi_{J(E)}$ of $J(E)$ and $\Phi_{J^*(E)}$
of $J^*(E)$ over $E$. Similarly, $M$-projectable vector fields $X_E$ on $E$
can be lifted to $E$-projectable vector fields $X_{J(E)}$ on $J(E)$ and $%
X_{J^*(E)}$ on $J^*(E)$.} \\[3mm]
Proof: These statements can all be inferred from the following formula,
which describes how a bundle automorphism $\Phi_E$ of $E$ over $M$ is lifted
to a bundle automorphism $\Phi_{J(E)}$ of $J(E)$ over $E$, namely simply by
taking the derivative. Indeed, we may think of a point $\, u_e \in J_e(E) \,$
as the jet or derivative $T_m \varphi$ of a local section $\varphi$ of $E$
at $m$ satisfying $\, e = \varphi(m)$, so in particular, $u_e$ is a linear
map from $T_m M$ to $T_e E$. Correspondingly, we may set
\begin{equation}
\Phi_{J(E)} \, u_e~=~T_e \Phi_E \circ u_e \circ \left( T_m \Phi_M
\right)^{-1}~.
\end{equation}
In local coordinates, the lifting of projectable vector fields is given by
\begin{equation}
X_{J(E)}~=~X^\mu \frac{\partial}{\partial x^\mu} \, + \, X^i \frac{\partial}{%
\partial q^i} \, - \, \left( \frac{\partial X^i}{\partial q^j} \, q_\mu^j \,
- \, \frac{\partial X^\nu}{\partial x^\mu} \, q_\nu^i \, + \, \frac{\partial
X^i}{\partial x^\mu} \right) \frac{\partial}{\partial q_\mu^i}~,
\end{equation}
and
\begin{eqnarray}
X_{J^*(E)} \!\! &=&\!\! X^\mu \frac{\partial}{\partial x^\mu} \, + \, X^i
\frac{\partial}{\partial q^i} \, - \left( \frac{\partial X^j}{\partial q^i}
\, p_j^\mu \, - \, \frac{\partial X^\mu}{\partial x^\nu} \, p_i^\nu \, + \,
\frac{\partial X^\nu}{\partial x^\nu} \, p_i^\mu \right) \frac{\partial}{%
\partial p_i^\mu}  \nonumber \\
& & \hspace{2.8cm} - \left( \frac{\partial X^i}{\partial x^\mu} \, p_i^\mu
\, + \, \frac{\partial X^\nu}{\partial x^\nu} \, p \right) \frac{\partial }{%
\partial p}~.
\end{eqnarray}
Just as in ordinary mechanics on cotangent bundles, one uses this lift to
define the {\em multimomentum map} \cite{marsden} which to each projectable
vector field $X_E$ on $E$ associates the $(n-1)$-form $J(X_E)$ on $J^*(E)$
defined by contraction with the canonical $n$-form $\theta$:
\begin{equation}
J(X_E)~=~i_{X_{J^*(E)}} \, \theta~.
\end{equation}
Now invariance of $\theta$ under bundle automorphisms of $J^*(E)$ that arise
from bundle automorphisms of $E$ by lifting implies that
\begin{equation}
L_{X_{J^*(E)}} \, \omega~=~0~,
\end{equation}
so that
\begin{equation}
i_{X_{J^*(E)}} \, \omega~=~d \, J(X_E)~,
\end{equation}
which means that $J(X_E)$ is a Hamiltonian $(n-1)$-form. In coordinates one
finds \cite{marsden}
\begin{equation}
J(X_E)~=~\left( p_i^\mu X^{i} \, + \, p \, X^\mu \right) d^n x_\mu \, - \, {%
\textstyle \frac{1}{2}} \left( p_i^\mu X^\nu \, - \, p_i^\nu X^\mu \right)
\, dq^i \wedge d^n x_{\mu\nu}  \label{noether}
\end{equation}
The first term on the right hand side of this equation, the only one present
in Kanatchikov's approach, corresponds to internal symmetry transformations,
whereas the remaining terms describe transformations (diffeomorphisms) that
act nontrivially on space-time; it is from this part of the multimomentum
map that one extracts the energy momentum tensor of field theory \cite%
{marsden2}.

\section{Hamiltonian forms of degree $n-1$ and their Poisson bracket}

In the previous section, we saw that the multimomentum map, which
encompasses the energy-momentum tensor as well as the Noether currents
associated with any kind of continuous symmetry in field theory, produces
Hamiltonian $(n-1)$-forms. The structure of all Hamiltonian $(n-1)$-forms is
completely described by the following \\[3mm]
{\bf Theorem \cite{forroe}:} {\em Hamiltonian $(n-1)$-forms on $J^*(E)$ are
the sum of three contributions:}

\begin{enumerate}
\item {\em the Noether current $J(X_E)$ associated to a projectable vector
field $X_E$ on $E$,}

\item {\em the pull-back of a horizontal $(n-1)$-form on $E$ to $J^*(E)$,}

\item {\em any closed $(n-1)$-form on $J^*(E)$.}
\end{enumerate}

In local coordinates, this decomposition (which is of course not unique) can
be written explicitly as follows. Let
\begin{equation}
X_E~=~X^\mu \frac{\partial}{\partial x^\mu} \, + \, X^i \frac{\partial}{%
\partial q^i}
\end{equation}
be a projectable vector field on $E$ and
\begin{equation}
f_0~=~f_0^\mu \, d^n x_\mu
\end{equation}
be an $(n-1)$-form on $J^*(E)$ obtained from a horizontal $(n-1)$-form on $E$
(with the same local coordinate expression) by pull-back: this means that
the coefficient functions $X^\mu$ depend only on the variables $x^\nu$ while
the coefficient functions $X^i$ and $f_0^\mu$ depend only on the variables $%
x^\nu$ and $q^j$. Define
\begin{equation}
f~=~J(X_E) \, + \, f_0~,
\end{equation}
so
\begin{equation}
f~=~\left( p_i^\mu X^i \, + \, p \>\! X^\mu \, + \, f_0^\mu \right) d^n
x_\mu \, - \, {\textstyle \frac{1}{2}} \left( p_i^\mu X^\nu \, - \, p_i^\nu
X^\mu \right) \, dq^i \wedge d^n x_{\mu\nu}~.
\end{equation}
Then $f$ is a Hamiltonian $(n-1)$-form, and the corresponding Hamiltonian
vector field $X_f$ reads
\begin{eqnarray}  \label{eq:VFCJETB3}
X_f \!\! &=&\!\! X^\mu \frac{\partial}{\partial x^\mu} \, + \, X^i \frac{%
\partial}{\partial q^i}  \nonumber \\
& &\!\! - \, \left( \frac{\partial X^j}{\partial q^i} \, p_j^\mu \, - \,
\frac{\partial X^\mu}{\partial x^\nu} \, p_i^\nu \, + \, \frac{\partial X^\nu%
}{\partial x^\nu} \, p_i^\mu \, + \, \frac{\partial X^\mu}{\partial q^i} \,
p \, + \, \frac{\partial f_0^\mu}{\partial q^i} \right) \frac{\partial}{%
\partial p_i^\mu} \\
& &\!\! - \, \left( \frac{\partial X^i}{\partial x^\mu} \, p_i^\mu \, + \,
\frac{\partial X^\nu}{\partial x^\nu} \, p \, + \, \frac{\partial f_0^\mu}{%
\partial x^\mu} \right) \frac{\partial}{\partial p}~.  \nonumber
\end{eqnarray}
We shall also write these expressions for $f$ and for $X_f$ in the form
\begin{equation}
f~=~f^\mu \, d^n x_\mu \, + \, {\textstyle \frac{1}{2}} \, f_i^{\mu\nu} \,
dq^i \wedge d^n x_{\mu\nu}~,
\end{equation}
and
\begin{equation}
X_f~=~\frac{\partial f^\mu}{\partial p} \, \frac{\partial}{\partial x^\mu}
\, + \, \frac{1}{n} \, \frac{\partial f^\mu}{\partial p_i^\mu} \, \frac{%
\partial}{\partial q^i} \, - \, \left( \frac{\partial f^\mu}{\partial q^i}
\, - \, \frac{\partial f_i^{\mu\nu}}{\partial x^\nu} \right) \frac{\partial
}{\partial p_i^\mu} \, - \, \frac{\partial f^\mu}{\partial x^\mu} \, \frac{%
\partial}{\partial p}~.
\end{equation}
The theorem claims that up to a closed $(n-1)$-form, $f$ is the most general
Hamiltonian $(n-1)$-form $f$ on $J^*(E)$. Note the integrability constraints
that express themselves through the dependence of the coefficient functions
on the variables $p_i^\mu$ and $p$, which is affine (linear plus constant).

For the definition of a Poisson bracket between Hamiltonian $(n-1)$-forms $f$
and $g$, the first idea would be to set
\begin{equation}
\left\{ f,g \right\}^{\prime}~=~i_{X_g} i_{X_f} \, \omega~,
\end{equation}
since this gives
\begin{equation}
\left[ X_f , X_g \right]~=~- \; X_{\{f,g\}^\prime}~.
\end{equation}
But this bracket satisfies Jacobi's identity only up to an exact term:
\begin{equation}
\left\{ f , \left\{ g , h \right\}^{\prime} \right\}^{\prime} \, + \,
\left\{ g , \left\{ h , f \right\}^{\prime} \right\}^{\prime} \, + \,
\left\{ h , \left\{ f , g \right\}^{\prime} \right\}^{\prime}~ =~d \left(
i_{X_f} i_{X_g} i_{X_h} \, \theta \right)~.
\end{equation}
This disease can be cured \cite{forroe} by adding a correction term, which
is a uniquely defined exact $(n-1)$-form in order to guarantee that, as
before,
\begin{equation}
\left[ X_f , X_g \right]~=~- \; X_{\{f,g\}}~.
\end{equation}
Explicitly,
\begin{equation}
\left\{ f , g \right\}~ =~i_{X_g} i_{X_f} \, \omega \, + \, d \left( i_{X_g}
f \, - \, i_{X_f} g \, - \, i_{X_g} i_{X_f} \, \theta \right)~.
\label{forpoiss}
\end{equation}
It can be checked \cite{forroe} that this new bracket does satisfy Jacobi's
identity and hence provides the space of Hamiltonian $(n-1)$-forms on $J^*(E)
$ with the structure of a Lie algebra. By an explicit calculation in local
coordinates using the above expressions for $f$, $X_f$ and analogous ones
for $g$, $X_g$, one finds
\begin{eqnarray}
\left\{ f , g \right\} \!\! &=&\!\! \left[ \frac{\partial X^\nu}{\partial
x^\nu} \, g^\mu \, - \, f^\mu \, \frac{\partial Y^\nu}{\partial x^\nu} \, +
\, \frac{\partial f^\mu}{\partial q^i} \, Y^i \, - \, X^i \, \frac{\partial
g^\mu}{\partial q^i} \right] d^n x_\mu  \nonumber \\
& & - \left[ \left( \frac{\partial X^\nu}{\partial q^i} \, g^\mu \, - \,
f^\mu \, \frac{\partial Y^\nu}{\partial q^i} \right) + \, p_i^\mu \left(
\frac{\partial X^\nu}{\partial x^\rho} \, Y^\rho \, - \, X^\rho \frac{%
\partial Y^\nu}{\partial x^\rho} \right) \right. \\
& & \quad - \left. \, p \left( \frac{\partial X^\nu}{\partial q^i} \, Y^\mu
\, - \, X^\mu \, \frac{\partial Y^\nu}{\partial q^i} \right) \right] dq^i
\wedge d^n x_{\mu\nu}  \nonumber
\end{eqnarray}
The calculation shows that the correction terms in the definition of the new
bracket lead to strong cancellations and greatly simplifies the final result.

In order to extend the new Poisson bracket to Hamiltonian forms of arbitrary
degree, two problems need to be solved.

\begin{itemize}
\item Equation (\ref{forpoiss}) has to be modified by introducing signs
depending  on the degrees of the Hamiltonian forms such that a graded
version of  Jacobi's identity still holds.

\item For forms $f$ of degree other than $n-1$, $df$ no longer determines
the  Hamiltonian multivector field $X_f$ uniquely. This ambiguity has to be
fixed in a consistent way.
\end{itemize}

These questions are presently under investigation.\\
{\bf Note added (20.7.2001):}
Meanwhile, these problems have been solved by M. Forger, C.
Paufler and H. R{\"o}mer. Defining Poisson $(n-r)$-forms $f$ as Hamiltonean
forms such that
\begin{equation}
i_X\omega=0\quad\Rightarrow\quad i_X f=0
\end{equation}
for all multivectorfields $X$ then the following bracket
\begin{eqnarray}
\left\{  f,g\right\}  &=&(-1)^{(p-1)(q-1)}L_{X_{g}}f-L_{X_{f}%
}g+(-1)^{q-1}L_{X_{g}\wedge X_{f}}\theta\\
&=&(-1)^{p}i_{X_{f}}i_{X_{g}}\Omega\\
&&\quad+d\left[  (-1)^{(p-1)(q-1)}
i_{X_{g}}f-i_{X_{f}}g+(-1)^{q-1}i_{X_{f}}i_{X_{g}}\theta\right]
\end{eqnarray}
is well defined and satisfies the graded Jacobi identity for
Poisson forms of arbitrary degree.

\end{document}